\documentclass[11pt]{article}

\usepackage{amssymb}
\usepackage[dvips]{graphicx}
\usepackage{citesort}

\begin{document}
\begin{titlepage}
\thispagestyle{empty}
\vspace*{1cm}
\begin{center}
{\LARGE\bf\sf  Modular bootstrap in Liouville field theory} \\
\end{center}

\begin{center}

\vspace*{2cm}

    {\large\bf\sf
    Leszek Hadasz${}^\dag$\footnote{\emph{e-mail}: hadasz@th.if.uj.edu.pl}$\!\!\!\!,\ \,$
    Zbigniew Jask\'{o}lski${}^\ddag$\footnote{\emph{e-mail}: jask@ift.uni.wroc.pl}
    and
    Paulina Suchanek${}^\ddag$\footnote{\emph{e-mail}: paulina@ift.uni.wroc.pl}
    }
     \\
\vskip 3mm
    ${}^\dag$ M. Smoluchowski Institute of Physics,
    Jagiellonian University \\
    Reymonta 4,
    30-059~Krak\'ow, Poland, \\

\vskip 3mm
    ${}^\ddag$ Institute of Theoretical Physics,
    University of Wroc{\l}aw \\
    pl. M. Borna, 50-204~Wroc{\l}aw, Poland. \\
\end{center}

\vspace*{1cm}
\begin{abstract}
The modular matrix for the  generic  1-point  conformal blocks on the torus is
expressed in terms of the fusion matrix for the 4-point blocks on the sphere.
The modular invariance of the toric 1-point functions
in the Liouville field theory with DOZZ structure constants is proved.
\end{abstract}

\vspace*{\fill}


\end{titlepage}
\section{Introduction}

The basic consistency conditions for any CFT on closed surfaces
are the crossing symmetry of the 4-point function on the sphere
and the modular invariance of the 1-point function on the torus \cite{Sonoda:1988fq}.
In the case of the Liouville field theory defined by the DOZZ structure constants
\cite{Dorn:1994xn,Zamolodchikov:1995aa} the first issue was addressed by Ponsot and Teschner \cite{Ponsot:1999uf,Ponsot:2000mt}.
They derived a system of functional equations
for the braiding and the fusion matrices %of the Liouville theory
and constructed its explicit solutions. The problem of crossing symmetry
in the Liouville field theory can than be reduced to a certain orthogonality relations satisfied by
the Barnes functions \cite{Kashaev}.
The exact form of the braiding and the fusion matrices
can be also
derived by direct calculations of the exchange relation of
chiral vertex operators in the
free field representation \cite{Teschner:2001rv,Teschner:2003en}
 (see also \cite{Gervais:1993fh} for an earlier construction).
Up to our knowledge the second consistency condition has not yet been analyzed in the Liouville field theory.

Although derived in the context of the Liouville field theory the results of  \cite{Ponsot:1999uf,Ponsot:2000mt} and
\cite{Teschner:2001rv,Teschner:2003en} are more universal.
From the point of view of the Moore--Seiberg approach \cite{Moore:1988qv} to classification of rational CFT models
the braiding and the fusion matrices found in \cite{Ponsot:1999uf,Ponsot:2000mt}
are two of the generators of the duality grupoid
describing the chiral structure of any CFT with the Liouville continuous spectrum.
The only missing generator is the modular matrix relating 1-point  conformal blocks  on tori
with modular parameters $\tau$ and $-{1\over \tau}$.

Our aim in the present paper is to derive an explicit form of the modular matrix in the case of Liouville spectrum
and to prove the modular invariance of the Liouville 1-point  functions on the torus.
The first result is based on recently discovered relations \cite{Poghossian:2009mk,Hadasz:2009db}
between 1-point conformal blocks on the torus and  4-point
conformal blocks on the sphere inspired by a corresponding relation between Liouville correlation
functions first proposed by  Fateev, Litvinov, Neveu and Onofri in \cite{Fateev:2009me}.
The second follows from the relation between DOZZ structure constants also suggested by the FLNO relation.

There are at least three problems which are natural continuation of
the present work. The first one is a more detailed analysis of the Liouville modular grupoid.
Since the Liouville  spectrum is continuous
the generators of the modular grupoid can be analytically continued
well outside the spectrum. For instance in the case of a degenerate
weight the integral over continuous spectrum localizes giving
rise to a finite dimensional fusion matrix \cite{Ponsot:2003ju,Hadasz:2004cm}.
The question arises whether any (irreducible) modular grupoid for Virasoro conformal blocks
can be obtained by an analytic continuation of the Liouville one.
The second  is to extend the results of the present work
to the $H_3^+$ WZNW model \cite{Ribault:2005wp,Hikida:2007tq}.
Finally, the third problem is to complete the verification of the consistency conditions
\cite{Lewellen:1991tb}
for the Liouville field theory
on bordered surfaces.

\section{Conformal blocks}

The 1-point toric and the 4-point spherical conformal blocks are defined by
\begin{eqnarray}
 \label{1tblock}
\mathcal{F}_{c, \Delta}^{\lambda}(q)
&=&
q^{\Delta -\frac{c}{24}} \,
\sum_{n =0}^{\infty} q^{n} \,
    F^{\lambda, n}_{c, \Delta},
\\
\label{1tblockcoef}
F^{\lambda, n}_{c, \Delta}
&=&
\sum_{n=|M|=|N|}  \rho(\nu_{\Delta,N}, \nu_{\lambda} , \nu_{\Delta,M})\,
\left[ B^n_{c,\Delta} \right]^{MN},
\end{eqnarray}
and
\begin{eqnarray}
\label{4sblock}
\mathcal{F}_{c,\Delta}
\left[^{\Delta_3 \; \Delta_2}_{\Delta_4 \; \Delta_1} \right](z)
&=&
z^{\Delta - \Delta_2 - \Delta_1} \left( 1 +
\sum_{n\in \mathbb{N}} z^n
    F^n_{c, \Delta}
    \left[^{\Delta_3 \; \Delta_2}_{\Delta_4 \; \Delta_1} \right] \right),
\\
\label{4sblockcoef}
F^{n}_{c, \Delta} \left[^{\Delta_3 \; \Delta_2}_{\Delta_4 \; \Delta_1} \right]
&=&
\sum_{n=|M|=|N|}\rho(\nu_4, \nu_3 ,\nu_{\Delta,M} )
  \left[ B^n_{c,\Delta} \right]^{MN}
  \rho(\nu_{\Delta,N}, \nu_2 , \nu_1 ),
\end{eqnarray}
respectively. $\rho(\xi_3,\xi_2,\xi_1)$ denotes the 3-point spherical conformal block and
{\small $\left[B^{n}_{c, \Delta}\right]^{MN}$} is the
inverse of the Gram matrix
\[
\left[B^{n}_{c, \Delta}\right]_{MN}
=
\left\langle\nu_{\Delta,N}\big|\nu_{\Delta,M}\right\rangle,
\hskip 1cm
|M| = |N| = n,
\]
calculated in the standard basis of the Verma module ${\cal V}_\Delta$:
$$
\nu_{\Delta,M}
 = L_{-M}\nu_{\Delta} \; \equiv \;
L_{-m_j}\ldots L_{-m_1}\nu_{\Delta}\,,
$$
with $M = \{m_1,m_2,\ldots,m_j\}\subset \mathbb{N}$ standing for an arbitrary ordered  set of  indices
$
m_j \leq \ldots \leq m_2 \leq m_1
$
and $\nu_{\Delta}\in \mathcal{V}_{\Delta}$  being the highest weight state.
In the case of torus the 1-point  elliptic conformal block $\mathcal{H}_{c, \Delta}^{\lambda}(\tilde q)$ is defined by:
\begin{eqnarray}
\label{1tell}
\mathcal{F}_{c, \Delta}^{\lambda}(\tilde q)&=&   \tilde q^{\Delta-\frac{c-1}{24}}\,\eta(\tilde q)^{-1}\,
 \mathcal{H}_{c, \Delta}^{\lambda}(\tilde q),
\end{eqnarray}
where the elliptic variable $\tilde q$ is related to the torus moduli parameter $\tau$ by
$
\tilde q={\rm e}^{2\pi i \tau}
$
and $\eta(\tilde q)$ is the Dedekind eta function.

The 4-point  elliptic conformal block on the sphere
$\mathcal{H}_{\Delta}\! \left[^{\Delta_3 \;\Delta_2}_{\Delta_4 \; \Delta_1} \right]\!(q)$ is given by \cite{Zamolodchikov:3}:
 \begin{eqnarray}
\label{4sell}
  \mathcal{F}_{\Delta}\! \left[^{\Delta_3 \;\Delta_2}_{\Delta_4 \; \Delta_1} \right] (z)
 & =&
(16q)^{\Delta - \frac{c-1}{24}}\ z^{\frac{c-1}{24} - \Delta_1 -  \Delta_2} \
 (1- z)^{\frac{c-1}{24} - \Delta_2 - \Delta_3}\
\\
\nonumber
 & \times &\theta_3^{\frac{c - 1}{2}
- 4 (\Delta_1 + \Delta_2 + \Delta_3 + \Delta_4) }  \
 \mathcal{H}_{\Delta}\! \left[^{\Delta_3 \;\Delta_2}_{\Delta_4 \; \Delta_1} \right]\!(q).
\end{eqnarray}
The variable $q$ is related to the moduli parameter $z$ of the 4-punctured sphere  by
\begin{eqnarray}
\label{tau}
q(z)= {\rm e}^{i \pi \tau}, \qquad  \tau(z) =i \frac{K(1-z)}{K(z)},
\end{eqnarray}
where $K(z)$ is the  complete elliptic integral of the first kind.

\section{Modular matrix}

Our starting point are the identities conjectured in \cite{Poghossian:2009mk} and proved in our previous paper \cite{Hadasz:2009db}:
\begin{eqnarray}
\label{rel:among:blocksI} \mathcal{H}_{c,
\Delta_\alpha}^{\lambda}\left(q^2\right) &=&
\mathcal{H}_{c',{\Delta'}_{\!\!\alpha'}}\!\left[^{{1\over
2b'}\;\;{\lambda\over \sqrt{2}}}_{{{1\over 2b'}\;\;\:{1\over
2b'}}}\right]\!(q) \;,\;\;\;\;\;b'=\textstyle {b\over
\sqrt{2}}\;,\;\;\;\;\;\alpha' = {\sqrt{2}\alpha }\, ,
\end{eqnarray}
and
\begin{eqnarray}
\label{rel:among:blocksII} \mathcal{H}_{c,
\Delta_\alpha}^{\lambda}\left(q^2\right) &=&
\mathcal{H}_{c',{\Delta'}_{\!\!\alpha'}}\!\left[^{{b'\over
2}\;\;{\lambda\over \sqrt{2}}}_{{{b'\over 2}\;\;\:{b'\over
2}}}\right]\!(q) \;,\;\;\;\;\;b'= \textstyle
{\sqrt{2}b}\;,\;\;\;\;\;\alpha' = {\sqrt{2}\alpha}\, ,
\end{eqnarray}
where
$$
\textstyle c=1+6\left(b+{1\over b} \right)^2,\;\;\;\;\;\;\Delta_\alpha = {1\over 4}\left(b+{1\over b} \right)^2 - {1\over 4}\alpha^2.
$$
Let us observe that the crossing
symmetry transformation $z\to 1-z$ on the sphere implies the modular transformation
%\begin{equation}\label{modtra}
$\tau \to -{1\over \tau}$
%\end{equation}
for $\tau(z)$ given by (\ref{tau}) and therefore the modular transformation of the elliptic  variable $\tilde q =q^2$ of the torus.
It follows that the crossing symmetry on the sphere on the r.h.s.\ of
(\ref{rel:among:blocksI}) and (\ref{rel:among:blocksII})
can be interpreted as the modular transformation of the toric 1-point function on the l.h.s.\ of these equations.
This yields the relation between
the modular matrix for the  1-point blocks on the torus defined by
\begin{eqnarray}
\label{mmod:def}
{\cal F}_{c,\Delta_s}^\lambda(q(\tau))
& = &
(-i\tau)^{-\Delta_\lambda}\int\limits_{i\mathbb{R}_+}
{\mathrm{d}\lambda_t  \over 2i}\;
{\sf S}_{\lambda_s\lambda_t}^{c, \lambda}\,
{\cal F}_{c,\Delta_t}^\lambda\left(q\left(\textstyle -{1\over \tau}\right)\right)
\end{eqnarray}
and
the fusion matrix
for the spherical 4-point blocks
\begin{eqnarray}
\label{fusion}
{\cal F}_{c,\Delta_s}\!\left[^{\Delta_3\:\Delta_2}_{\Delta_4\:\Delta_1}\right]\!(z)
& = &
 \int\limits_{i\mathbb{R}_+}
{\mathrm{d}\lambda_t  \over 2i}\;
{\sf F}^c_{\lambda_s\lambda_t}\!\left[^{\lambda_3\:\lambda_2}_{\lambda_4\:\lambda_1}\right]
{\cal F}_{c,\Delta_t}\!\left[^{\Delta_1\:\Delta_2}_{\Delta_4\:\Delta_3}\right]\!(1-z)\, .
\end{eqnarray}
Indeed using equations (\ref{1tell}, \ref{4sell}, \ref{rel:among:blocksI}, \ref{fusion}) and the relations:
\begin{eqnarray}
\label{etaandtheta}
\eta\!\left(e^{- \frac{2 \pi i}{\tau}}\right) &=& {\sqrt{-i \tau}}\;\eta\!\left(e^{2 \pi i \tau} \right),
\;\;\;\;\;
\theta_3\!\!\left(e^{- \frac{ \pi i}{\tau}}\right)\;=\;{\sqrt{-i \tau}}\;\theta_3\!\!\left(e^{ \pi i \tau} \right),
%\\ 3{Q'}^2 - 4\Delta'_{\lambda'}-12\Delta'_{1\over 2b'} & = & -2\Delta_\lambda -1
\end{eqnarray}
one gets:
\begin{equation}
\label{mod:fusI}
{\sf S}_{\lambda_s\lambda_t}^{c, \lambda}
=
2^{2(\lambda_s^2-\lambda_t^2)+{1\over 2}}\
{\sf F}^{c'}_{\sqrt{2}\lambda_s\,\sqrt2\lambda_t}\!
\left[^{{1\over
2b'}\;\;{\lambda\over \sqrt{2}}}_{{{1\over 2b'}\;\;\:{1\over
2b'}}}\right] ,\;\;\;\;\;\;b'=\textstyle {b\over
\sqrt{2}}\ ,
\end{equation}
or (using the relation (\ref{rel:among:blocksII}))
\begin{equation}
\label{mod:fusII}
{\sf S}_{\lambda_s\lambda_t}^{c, \lambda}
=
2^{2(\lambda_s^2-\lambda_t^2)+{1\over 2}}\
{\sf F}^{c'}_{\sqrt{2}\lambda_s\,\sqrt2\lambda_t}\!
\left[^{{b'\over 2}\;\;{\lambda\over \sqrt{2}}}_{{{b'\over 2}\;\;\:{b'\over 2}}}\right] ,\;\;\;\;\;\;b'=\textstyle {
\sqrt{2}b}\ .
\end{equation}
Some remarks concerning the application of formula (\ref{fusion}) in the derivation above are in order.
Let us consider the fusion matrix  for
$\lambda_1 = \lambda_3 = \lambda_4=\eta,\  \lambda_2=\lambda$.
In the present parametrization of conformal weights it reads\footnote{Definitions and discussion of some basic properties of the
functions $\Gamma_b$ and $S_b$ and $\Upsilon_b$ appearing below can be
found in \cite{Ponsot:1999uf,Ponsot:2000mt}; see also the Appendix. For the
detailed discussion of the Barnes special functions the reader may
consult the papers \cite{Barnes,Shintani}.}
 \cite{Ponsot:1999uf,Ponsot:2000mt}:
\begin{eqnarray}
\label{fusion:matrix}
&&
\hskip -1cm
{\sf F}^c_{\lambda_s \lambda_t}\!\left[^{\eta\:\lambda}_{\eta\:\eta}\right]
=
\frac{
\Gamma_b(\frac{Q}{2} - \eta - \frac{\lambda_t}{2})
\Gamma_b(\frac{Q}{2} - \eta + \frac{\lambda_t}{2})
\Gamma_b(\frac{Q}{2} - \frac{\lambda_t}{2})
\Gamma_b(\frac{Q}{2} + \frac{\lambda_t}{2})
}
{
\Gamma_b(\frac{Q}{2} - \eta - \frac{\lambda_s}{2})
\Gamma_b(\frac{Q}{2} - \eta + \frac{\lambda_s}{2})
\Gamma_b(\frac{Q}{2} - \frac{\lambda_s}{2})
\Gamma_b(\frac{Q}{2} + \frac{\lambda_s}{2})
}
\\[4pt]
\nonumber
& \times &
\frac{
\Gamma_b(\frac{Q}{2} - \frac{\lambda}{2} - \frac{\eta}{2} - \frac{\lambda_t}{2})
\Gamma_b(\frac{Q}{2} + \frac{\lambda}{2} - \frac{\eta}{2} - \frac{\lambda_t}{2})
\Gamma_b(\frac{Q}{2} - \frac{\lambda}{2} - \frac{\eta}{2} + \frac{\lambda_t}{2})
\Gamma_b(\frac{Q}{2} + \frac{\lambda}{2} - \frac{\eta}{2} + \frac{\lambda_t}{2})
}
{
\Gamma_b(\frac{Q}{2} - \frac{\lambda}{2} - \frac{\eta}{2} - \frac{\lambda_s}{2})
\Gamma_b(\frac{Q}{2} + \frac{\lambda}{2} - \frac{\eta}{2} - \frac{\lambda_s}{2})
\Gamma_b(\frac{Q}{2} - \frac{\lambda}{2} - \frac{\eta}{2} + \frac{\lambda_s}{2})
\Gamma_b(\frac{Q}{2} + \frac{\lambda}{2} - \frac{\eta}{2} + \frac{\lambda_s}{2})
}
\\[4pt]
\nonumber
& \times &
\frac{\Gamma_b(Q+\lambda_s)\Gamma_b(Q-\lambda_s)}{\Gamma_b(\lambda_t)\Gamma_b(-\lambda_t)}\
I^c_{\lambda_s\lambda_t}\!\left[^{\eta\:\lambda}_{\eta\:\eta}\right]
\end{eqnarray}
where
\begin{eqnarray}
\label{I:c}
\nonumber
I^c_{\lambda_s\lambda_t}\!\left[^{\eta\:\lambda}_{\eta\:\eta}\right]
& = &
\int\limits_{i{\mathbb R}}\frac{{\mathrm d}\tau}{i}
\left[
\frac{
S_b(\frac{Q}{2} - \frac{\lambda}{2} + \tau)
S_b(\frac{Q}{2} + \frac{\lambda}{2} + \tau)
}
{
S_b(Q-\frac{\lambda_s}{2} + \frac{\eta}{2} + \tau-0^+)
S_b(Q+\frac{\lambda_s}{2} + \frac{\eta}{2} + \tau-0^+)
}
\right.
\\[-6pt]
\\[-6pt]
\nonumber
&&
\left.
\hskip 1cm
 \times \ \frac{
S_b(\frac{Q}{2} - \frac{\eta}{2} + \tau)
S_b(\frac{Q}{2} + \frac{\eta}{2} + \tau)
}
{
S_b(Q-\frac{\lambda_t}{2} - \frac{\eta}{2} + \tau-0^+)
S_b(Q+\frac{\lambda_t}{2} - \frac{\eta}{2} + \tau-0^+)
}
\right].
\end{eqnarray}
The relations (\ref{fusion}), (\ref{fusion:matrix}) and  (\ref{I:c}) were derived for conformal
weights from the spectrum of the Liouville field theory, $\lambda_s,\lambda_t,\lambda,\eta\in i\mathbb{R},$
while in our derivation  analytic continuations to $\eta = {1\over 2b}$ and  $\eta = {b\over 2}$ are required.

Let us start with the analytic continuation of
\(
I^c_{\lambda_s\lambda_t}\!\left[^{\eta\:\lambda}_{\eta\:\eta}\right]
\).
For $\lambda_s, \lambda_t,\lambda,\eta \in i{\mathbb R}$ the integrand in (\ref{I:c}) has poles
(coming from the poles of the functions $S_b$ in the numerator) located at $\Re\,\tau < 0$
(to the left from the integration contour)
and poles coming from the zeroes of the $S_b$ functions in the denominator, located
at $\Re\,\tau > 0$ (to the right from the integration contour). Some of these
poles move when we analytically continue in $\eta$. If they cross the imaginary axis
the process of analytic continuation requires an appropriate smooth deformation of the contour
of integration in (\ref{I:c}).
 As was
discussed in \cite{Ponsot:2000mt} such deformation is possible unless
there are some poles with locations coinciding at the terminal value of $\eta,$ which
``pinch'' the $\tau$ integration contour in between. This happens for instance when the terminal value
of $\eta$ corresponds to a degenerate weight, $\eta = {mb} + \frac{n}{b},\ m,n \in {\mathbb N},$ but neither
for $\eta = \frac{1}{2b}$ nor for $\eta = \frac{b}{2}$.
Thus
\(
I^c_{\lambda_s\lambda_t}\!\left[^{\eta\:\lambda}_{\eta\:\eta}\right]
\)
remains regular for $\lambda_t\in i{\mathbb R}$ while $\eta \to \frac{1}{2b}$ or $\eta \to \frac{b}{2}$.

The product of $\Gamma_b$ functions appearing in (\ref{fusion:matrix}) has poles moving with $\eta$
on both sides of the contour:
\begin{eqnarray*}
{\lambda_t} &=& \pm 2\left(\frac{Q}{2} - \eta + mb + nb^{-1}\right),
\\
{\lambda_t} &=& \pm 2\left(\frac{Q}{2} -\frac{\eta}{2} -\frac{\lambda}{2} + mb + nb^{-1}\right),
\;\;\;\;\;
{\lambda_t} \;=\;\pm 2\left(\frac{Q}{2} -\frac{\eta}{2} +\frac{\lambda}{2} + mb + nb^{-1}\right).
\end{eqnarray*}
For $\eta \to \frac{1}{2b}$ and for $\eta \to \frac{b}{2}$
none of these poles crosses the imaginary axis. Thus the analytic continuation
of the fusion formula (\ref{fusion}) from imaginary values $\eta \in  i\mathbb R$ to
$\eta = \frac{1}{2b}$ or to $\eta = \frac{b}{2}$ does not change
the integration contour. This justifies our definition of the modular matrix (\ref{mmod:def}).
It also implies that the fusion matrices on the right hand side of equations (\ref{mod:fusI}), (\ref{mod:fusII})
are just analytic continuation of the fusion matrices from the Liouville physical weights to
$\eta = \frac{1}{2b}$ and to $\eta = \frac{b}{2}$.

Let us finally note that parallel reasoning
with respect to the $\lambda_s$ variable shows that  the fusion
matrix ${\sf F}^c_{\lambda_s \lambda_t}\!\left[^{\eta\:\lambda}_{\eta\:\eta}\right]$,
 multiplied by its conjugation and integrated over $\lambda_s$ enjoys
the usual orthogonality properties, which ensure the crossing
symmetry
\begin{equation}
\label{crossingsym}
\left\langle \phi_{\eta} \phi_{\eta}
\phi_{\lambda}(z) \phi_{\eta}  \right\rangle^{\!c}
=\left\langle \phi_{\eta} \phi_{\eta}
\phi_{\lambda}(1-z) \phi_{\eta}  \right\rangle^{\!c}
\end{equation}
of the corresponding four-point Liouville correlation
function:
\begin{eqnarray}
\label{4pointfunction} \left\langle \phi_{\eta} \phi_{\eta}
\phi_{\lambda}(z) \phi_{\eta}  \right\rangle^{\hspace{-1pt}c}
&=& \left|\left(
z(1-z)\right)^{-\frac{Q^2}{4}+ \frac{\eta^2}{4}+
\frac{\lambda^2}{4}}
(\theta_3(q))^{-Q^2+3\eta^2+\lambda^2}\right|^2 \,
\\
&\times&
 \int_{i\mathbb{R}^+} \frac{\mathrm{d}\lambda_s}{2i} \left|  (16q)^{-{\lambda^2_s \over 4}}
H_{c,\Delta}\left[^{ \eta \;  \lambda}_{\eta \;\eta}
\right](q)\right|^2 \,
 C_{c}(-\eta,\eta,\lambda_s) C_c(-\lambda_s,\lambda,\eta).
 \nonumber
\end{eqnarray}

\section{Modular invariance}

In this section we shall prove that for $\lambda \in i\mathbb{R}$ the Liouville 1-point functions on the torus
satisfy the modular invariance condition \cite{Sonoda:1988fq}:
\begin{equation}
\label{modinv}
\langle \phi_{\lambda} \rangle_{-{1\over \tau}} =
|\tau |^{2 \Delta_\lambda}
\langle \phi_{\lambda} \rangle_\tau\ .
\end{equation}
In the  Liouville field theory the  1-point function can be expressed in terms of the elliptic blocks
 as follows:
\begin{equation}
\label{1pointfunction}
\langle \phi_{\lambda} \rangle_{\tau} = \int\limits_{i\mathbb{R}^+}\!
{\mathrm{d}\lambda_s  \over 2i}\left| \tilde q^{- \frac{\lambda_s^2}{4}} \, \eta(\tilde q)^{-1}
\mathcal{H}_{c, \Delta_s}^{\lambda}(\tilde q) \right|^2 \,
C_c\left(-\lambda_s,\lambda,\lambda_s \right),
\end{equation}
 where $\tilde q=q^2 ={\rm e}^{2\pi i \tau}$ and the DOZZ structure constants are given by:
\begin{eqnarray*}
C_c\left(\lambda_1,\lambda_2,\lambda_3 \right)
&=& \left[ \pi \mu \gamma(b^2) b^{2-2b^2}\right]^{-\frac{1}{2b}(\lambda_3 +\lambda_2 +\lambda_1+ Q)}
 \\
 &\times&
 \!\!\frac{\Upsilon_{b}(b)
\Upsilon_{b}(Q+\lambda_3)\Upsilon_{b}(Q+\lambda_2)\Upsilon_{b}(Q+\lambda_1)}
{\Upsilon_{b}\left(\frac{Q+\lambda_3 +\lambda_2 +\lambda_1}{2}  \right)
 \Upsilon_{b}\left(\frac{Q+\lambda_3 +\lambda_2 -\lambda_1}{2}\right)
 \Upsilon_{b}\left(\frac{Q+\lambda_3 -\lambda_2 +\lambda_1}{2}\right)
 \Upsilon_{b}\left(\frac{Q-\lambda_3 +\lambda_2 +\lambda_1}{2} \right)
}\,.
\end{eqnarray*}
Using the explicit form of the modular matrix  for the Liouville spectrum $\lambda, \lambda_s \in i \mathbb{R}$
(\ref{mod:fusI})
 one could in principle
analyze the behavior of the 1-loop function by direct calculations.

There is however a simpler derivation suggested by
the relation between the 1-point Liouville function on the torus (\ref{1pointfunction}) and the
4-point Liouville function on the sphere (\ref{4pointfunction})
first proposed by  Fateev, Litvinov, Neveu and Onofri in \cite{Fateev:2009me}.
It should be stressed that the FLNO relation
was the original inspiration for relations between conformal blocks (\ref{rel:among:blocksI}), (\ref{rel:among:blocksII})
 \cite{Poghossian:2009mk,Hadasz:2009db}.
So it was for the following relations between the Liouville structure constants:
\begin{eqnarray}
\label{strcon1}
C_c(-\lambda_s,\lambda,\lambda_s)
=
16^{-\lambda_s^2} g_1(\lambda,b)
\textstyle\;
C_{c'}\!\left(-{1\over 2b'},{1\over 2b'},{\sqrt{2}\lambda_s} \right)
C_{c'}\!\left(-{\sqrt{2}\lambda_s } ,{\lambda\over \sqrt{2}} ,{1\over 2b'} \right)
,\;\;b'={b\over \sqrt{2}}\,,
\\[6pt]
\label{strcon2}
C_c(-\lambda_s,\lambda,\lambda_s)
=
16^{-\lambda_s^2} g_{2}(\lambda,b)
\textstyle\;
C_{c'}\!\left(-{b'\over 2},{b'\over 2},{\sqrt{2}\lambda_s } \right)
C_{c'}\!\left(-{\sqrt{2}\lambda_s} ,{\lambda\over \sqrt{2}} ,{b'\over 2} \right)
,\;\;b'={\sqrt{2}b}\,,
\end{eqnarray}
where:
\begin{eqnarray*}
\label{g1}
g_1(\lambda,b) &=&
 \left[\pi \mu \gamma(b^2) b^{2-2b^2}\right]^{-\frac{1}{b}\left(\frac{Q}{2}+  \frac{\lambda}{2}\right)}
 \left[\pi \mu \gamma(b'^2) b'^{2-2b'^2}\right]^{\frac{1}{b'}(Q'+\frac{1}{4b'} +  \frac{\lambda}{2\sqrt{2}})}
 \\[6pt] \nonumber
 &\times&
2^{{b^2\over 2} + \frac{2}{b^2}-\frac34 + \frac{3b}{4} \lambda + \frac{1}{2b}\lambda + {1\over 2}\lambda^2 }
b^{6-{4\over b^2}} \gamma^{-2}(b^{-2}) \,
\frac{\Upsilon_{b}( \frac{b}{2}) }{\Upsilon_{b}(b)} \,
 \,
 \frac{\Upsilon_{b}(\frac{1}{2b}- \frac{\lambda}{2})}
{\Upsilon_{b}(\frac{Q}{2}+ \frac{\lambda}{2})}\ ,\;\;\;\;\;\;b'={b\over \sqrt{2}}\,,
\\[6pt]
\label{g2}
g_2(\lambda,b) &=&
 \left[\pi \mu \gamma(b^2) b^{2-2b^2}\right]^{-\frac{1}{b}\left(\frac{Q}{2}+  \frac{\lambda}{2}\right)}
 \left[\pi \mu \gamma(b'^2) b'^{2-2b'^2}\right]^{\frac{1}{b'}(Q'+\frac{b'}{4} +  \frac{\lambda}{2\sqrt{2}})}
 \\ \nonumber
 &\times&
2^{2b^2 + \frac{1}{2b^2}-\frac34 + \frac{3}{4b} \lambda + \frac{b}{2}\lambda + {\lambda^2\over 2} }
b^{4b^2-6} \gamma^{-2}(b^2) \,
\frac{\Upsilon_{b}( \frac{1}{2b}) }{\Upsilon_{b}(\frac{1}{b})} \,
 \frac{\Upsilon_{b}(\frac{b}{2}- \frac{\lambda}{2})}
{\Upsilon_{b}(\frac{Q}{2}+\frac{\lambda}{2})}\ ,\;\;\;\;\;\;b'={\sqrt{2}b}\,.
\end{eqnarray*}
The relations above can be obtained using
the following identities for the $\Upsilon$-function
 \cite{Fateev:2009me}:
\begin{eqnarray}
\label{Upsilon:relations}
\nonumber
\Upsilon_b(2x)
& = &
2^{4\left(x-Q/4\right)^2}\
\frac{
\Upsilon_b(x)\Upsilon_b\left(x+{\textstyle \frac12}b\right)
\Upsilon_b\left(x+{\textstyle \frac12}b^{-1}\right)\Upsilon_b\left(x+{\textstyle \frac{1}{2}}Q\right)
}
{
\Upsilon_b^2\left({\textstyle \frac{1}{4}Q}\right)\Upsilon_b^2\left({\textstyle \frac{1}{4}}Q+{\textstyle \frac12}b\right)
}
\\[4pt]
\nonumber
& = &
2^{4x\left(x-\frac12 Q\right)+1}\
\frac{
\Upsilon_b(x)\Upsilon_b\left(x+{\textstyle \frac12}b\right)
\Upsilon_b\left(x+{\textstyle \frac12}b^{-1}\right)\Upsilon_b\left(x+{\textstyle \frac{1}{2}}Q\right)
}
{
\Upsilon_b\left({\textstyle \frac12}b\right)\Upsilon_b\left({\textstyle \frac12}b^{-1}\right)
},
\\[-6pt]
\\[-6pt]
\nonumber
\Upsilon_{\frac{b}{\sqrt2}}(x\sqrt{2})
&=&
2^{x\left(x-\frac{1}{b}-\frac12 b\right)+\frac12}\
\Upsilon_{\frac{b}{\sqrt2}}\left({\textstyle\frac{b}{\sqrt 2}}\right)
\frac{\Upsilon_b(x)\Upsilon_b\left(x+{\textstyle\frac{1}{2}}b\right)}{\Upsilon_b\left({\textstyle\frac{1}{2}}b\right)\Upsilon_b(b)},
\\[4pt]
\nonumber
\Upsilon_{b\sqrt 2}(x\sqrt{2})
&=&
2^{x\left(x-\frac{1}{2b}- b\right)+\frac12}\
\Upsilon_{b\sqrt 2}\left({\textstyle\frac{b^{-1}}{\sqrt 2}}\right)
\frac{\Upsilon_b(x)\Upsilon_b\left(x+{\textstyle\frac{1}{2}}b^{-1}\right)}
{\Upsilon_b\left({\textstyle\frac{1}{2}}b^{-1}\right)\Upsilon_b(b^{-1})}.
\end{eqnarray}
For completeness we present a derivation of these formulae in the Appendix.
Relations (\ref{rel:among:blocksI}) and (\ref{strcon1}) imply:
\begin{equation}
\label{FLNO1}
 \left\langle \phi_{\lambda} \right\rangle^c_{\tau}
 =
 f(\lambda, q, b)\, g_1(\lambda,b) \,
\left\langle \phi_{\frac{1}{2b'}} \phi_{\frac{1}{2b'}}
\phi_{\frac{\lambda}{\sqrt2}}(z) \phi_{\frac{1}{2b'}}
\right\rangle^{\hspace{-3pt}c'}\;\;\;,\;\;\;b'={b\over \sqrt{2}}\,,
\end{equation}
while (\ref{rel:among:blocksII}) and (\ref{strcon2}) yield:
\begin{equation}
\label{FLNO2}
 \left\langle \phi_{\lambda} \right\rangle^c_{\tau}
 =
 f(\lambda,q, b^{-1})\, g_2(\lambda,b) \,
\left\langle \phi_{\frac{b'}{2}} \phi_{\frac{b'}{2}}
\phi_{\frac{\lambda}{\sqrt2}}(z) \phi_{\frac{b'}{2}}
\right\rangle^{\hspace{-3pt}c'}\;\;\;,\;\;\;b'={\sqrt{2}b}\,,
\end{equation}
where
\begin{eqnarray}
\nonumber
\label{fqb}
f(\lambda, q, b)
&=&
\left|\eta(q^2) \, \left( z(1-z)\right)^{-\frac{ b^2}{8}- \frac{3}{8 b^2} -\frac{1}{2} + \frac{\lambda^2}{8}}
(\theta_3(q))^{-\frac{b^2}{2}- \frac{1}{2 b^2} - 2 + \frac{\lambda^2}{2}}\right|^{-2} \,
\\[-6pt]
\\[-6pt]
\nonumber
&=&
\left|\eta(q^2) \, \left( \theta_2(q) \theta_4(q) \right)^{-\frac{b^2}{2}- \frac{3}{2 b^2} -2 + \frac{\lambda^2}{2}}
(\theta_3(q))^{ b^{-2}}\right|^{-2}.
\end{eqnarray}
Note that  (\ref{FLNO2}) is the original FLNO relation of \cite{Fateev:2009me}. Formulae (\ref{rel:among:blocksII})
and (\ref{strcon2}) provide a simple proof of this relation.
Relation (\ref{FLNO1}) is new but  of the same origin.

Using (\ref{FLNO1}) and the crossing symmetry of the 4-point function (\ref{crossingsym})
one can reduce the modular symmetry condition (\ref{modinv})
to the
relation
$$
f(\lambda, {\rm e}^{-i\pi {1\over \tau}}, b) = |\tau|^{\frac{b^2}{2} + \frac{1}{2 b^2} + 1 - \frac{\lambda^2}{2}} \,
f(\lambda, {\rm e}^{i\pi { \tau}}, b)
$$
which can be easily verified using formulae (\ref{etaandtheta}). This completes our proof of the modular invariance
in the Liouville field theory.

\setcounter{equation}{0}
\section*{Acknowledgements}
This work  was  supported by the Polish State Research
Committee (KBN) grant no. N N202 0859 33.
The work of L.H. was also supported by MNII grant 189/6.PRUE/2007/7.

\appendix
\section{Some identities satisfied by the Barnes functions}
\renewcommand{\theequation}{A.\arabic{equation}}
\setcounter{equation}{0}
For $\Re\, s >2$ the Barnes double zeta function can be defined as
\begin{eqnarray}
\label{zeta:definition}
\zeta_b(x;s)
& = &
\sum\limits_{n,m=0}^{\infty}
\left(x + mb + nb^{-1}\right)^{-s}.
\end{eqnarray}
Let us denote:
\begin{eqnarray}
\nonumber
\label{F:def}
F(b,x;s)
& = &
\zeta_b(x;s) - \zeta_b(Q/2;s).
\end{eqnarray}
With a help of the Mellin transform
\(
a^{-s} = \frac{1}{\Gamma(s)}\int_{0}^{\infty}\!dt\
t^{s-1}{\rm e}^{-at}
\)
we get:
\begin{eqnarray*}
F(b,x;s)
& = &
\Gamma(1-s)
\int_{\cal C}\,\frac{dt}{2\pi it}\
(-t)^{s}\ \frac{{\rm e}^{-tx}-{\rm e}^{-tQ/2}}{\left(1-{\rm e}^{-tb}\right)\left(1-{\rm e}^{-t/b}\right)}\,,
\end{eqnarray*}
where the integration contour $\cal C$ surrounds (in the positive direction) the cut of the $(-t)^s$ function
which is chosen along the positive real semi-axis. The last expression is valid also for $\Re\,s < 2.$
Since
\begin{eqnarray*}
\Gamma(1-s)\int_{\cal C}\,\frac{dt}{2\pi it}\
(-t)^{s}\ {\rm e}^{-t}
& = & 1,
\hskip 1cm
\int_{\cal C}\,\frac{dt}{2\pi it}\
(-t)^{s-1}
= 0,
\end{eqnarray*}
(the last formula holds for $\Re\, s < 1$) one has:
\begin{eqnarray}
\label{F:anal:cont}
&&
\hskip -1.5cm
F(b,x;s)
\; = \;
{\textstyle\frac12}\left({\textstyle\frac{Q}{2}}-x\right)^2
\\
\nonumber
& + &
\Gamma(1-s)\int_{\cal C}\,\frac{dt}{2\pi it }(-t)^s
\left[
\frac{{\rm e}^{-tx} -{\rm e}^{-tQ/2}}{\left(1-{\rm e}^{-tb}\right)\left(1-{\rm e}^{-t/b}\right)} -
\frac{Q/2 -x}{t} -
\frac12\left(Q/2-x\right)^2 {\rm e}^{-t} \right].
\end{eqnarray}
Formula (\ref{F:anal:cont}) is valid also for $s$ close to 0 and gives:
\begin{eqnarray}
\label{F:cont:val}
\nonumber
F(b,x;0) & = & {\textstyle\frac12}\left({\textstyle\frac{Q}{2}}-x\right)^2
\end{eqnarray}
together with
\begin{eqnarray}
\label{F:cont:der}
\log\Gamma_b(x)
& = &
\frac{\partial}{\partial s}F(b,x;s)\Big|_{s = 0}
\\
\nonumber
& = &
\int\limits_{0}^{\infty}\!\frac{dt}{t}
\left[
\frac{{\rm e}^{-tx} -{\rm e}^{-tQ/2}}{\left(1-{\rm e}^{-tb}\right)\left(1-{\rm e}^{-t/b}\right)}
-
\frac{Q/2 -x}{t}
-
\frac12\left(Q/2-t\right)^2 {\rm e}^{-t}
\right].
\end{eqnarray}

Separating the sum over integer $m$ and $n,$ appearing in the definition (\ref{zeta:definition})
of the Barnes zeta, onto sum of even $m,n,$ even $m$ and odd $n,$ odd $m$ and even $n$
and odd $m,n$ one gets:
\begin{eqnarray*}
\zeta_b(2x;s)
& = &
2^{-s}
\Big[
\zeta_b\left(x;s\right)
+
\zeta_b\left(x + {\textstyle \frac12}b;s\right)
+
\zeta_b\left(x + {\textstyle \frac12}b^{-1};s\right)
+
\zeta_b\left(x + {\textstyle \frac12}Q;s\right)
\Big]
\end{eqnarray*}
and similarly
\begin{eqnarray*}
\zeta_b({\textstyle \frac12}Q;s)
& = &
2^{-s}
\Big[
\zeta_b\left({\textstyle \frac14}Q;s\right)
+
\zeta_b\left({\textstyle \frac14}Q + {\textstyle \frac12}b;s\right)
+
\zeta_b\left({\textstyle \frac14}Q + {\textstyle \frac12}b^{-1};s\right)
+
\zeta_b\left({\textstyle \frac34}Q;s\right)
\Big].
\end{eqnarray*}
This gives:
\begin{eqnarray}
\label{F:2x}
\nonumber
F(b,2x;s)
& = &
2^{-s}
\Big\{
F\left(b,x;s\right) + F\left(b,x+{\textstyle \frac12}b;s\right)
+
F\left(b,x+{\textstyle \frac12}b^{-1};s\right) + F\left(b,x+{\textstyle \frac{Q}{2}};s\right)
\\[4pt]
\nonumber
& - &
F\left(b,{\textstyle \frac{Q}{4}};s\right) - F\left(b,{\textstyle \frac{Q}{4}}+{\textstyle \frac12}b;s\right)
-
F\left(b,{\textstyle \frac{Q}{4}}+{\textstyle \frac12}b^{-1};s\right) - F\left(b,{\textstyle \frac{3}{4}}Q;s\right)
\Big\}
\end{eqnarray}
and
\begin{eqnarray}
\label{Gamma:2:x}
\Gamma_b(2x)
& = & \exp\left\{\frac{\partial}{\partial s}F(b,2x;s)\Big|_{s=0}\right\}
\\
\nonumber
& = &
2^{-2\left(x-{\textstyle \frac{1}{4}}Q\right)^2}
\Upsilon_b\left({\textstyle \frac{1}{4}Q}\right)\Upsilon_b\left({\textstyle \frac{1}{4}}Q+{\textstyle \frac12}b\right)
\Gamma_b(x)\Gamma_b\left(x+{\textstyle \frac12}b\right)\Gamma_b
\left(x+{\textstyle \frac12}b^{-1}\right)\Gamma_b\left(x+{\textstyle \frac{1}{2}}Q\right).
\end{eqnarray}
Equation (\ref{Gamma:2:x}) and the definition
\(
\Upsilon_b^{-1}(x) =
\Gamma_b(x)\Gamma_b(Q-x)
\)
yield
\begin{eqnarray}
\label{Upsilon:2:x}
\Upsilon_b(2x)
& = &
2^{4\left(x-Q/4\right)^2}\
\frac{
\Upsilon_b(x)\Upsilon_b\left(x+{\textstyle \frac12}b\right)
\Upsilon_b\left(x+{\textstyle \frac12}b^{-1}\right)\Upsilon_b\left(x+{\textstyle \frac{1}{2}}Q\right)
}
{
\Upsilon_b^2\left({\textstyle \frac{1}{4}Q}\right)\Upsilon_b^2\left({\textstyle \frac{1}{4}}Q+{\textstyle \frac12}b\right)
}\ .
\end{eqnarray}
For $x = \frac12 b$ the formula (\ref{Upsilon:2:x}) gives:
\[
\Upsilon_b^2\left({\textstyle \frac{1}{4}Q}\right)\Upsilon_b^2\left({\textstyle \frac{1}{4}}Q+{\textstyle \frac12}b\right)
 =
 2^{\frac12\left(b^{-1} -b\right)^2}\Upsilon_b\left({\textstyle \frac12}b\right)\Upsilon_b\left({\textstyle \frac12}b^{-1}\right).
\]
Substituting this expression into (\ref{Upsilon:2:x}) we arrive at the double argument formula of FLNO
\begin{eqnarray}
\label{FLNO:double:argument}
\Upsilon_b(2x)
& = &
2^{4x\left(x-\frac12 Q\right)+1}\
\frac{
\Upsilon_b(x)\Upsilon_b\left(x+{\textstyle \frac12}b\right)
\Upsilon_b\left(x+{\textstyle \frac12}b^{-1}\right)\Upsilon_b\left(x+{\textstyle \frac{1}{2}}Q\right)
}
{
\Upsilon_b\left({\textstyle \frac12}b\right)\Upsilon_b\left({\textstyle \frac12}b^{-1}\right)
}\ .
\end{eqnarray}

Proceeding in a similar way and splitting the sum over $m$ appearing in (\ref{zeta:definition}) onto even and odd integers one gets:
\[
F\!\left({\textstyle\frac{b}{\sqrt 2}},x\sqrt 2;s\right)
-
F\!\left({\textstyle\frac{b}{\sqrt 2}},{\textstyle\frac{b}{\sqrt 2}};s\right)
=
2^{-\frac{s}{2}}
\Big\{
F\left(b,x;s\right) +F\left(b,x+{\textstyle\frac{b}{2}};s\right) - F\left(b,{\textstyle\frac{b}{2}};s\right) - F(b,b;s)
\Big\}
\]
and therefore:
\begin{equation}
\label{Gamma:scaling}
\Gamma_{\frac{b}{\sqrt2}}(x\sqrt{2})
=
2^{-\frac12x\left(x-\frac{1}{b}-\frac12 b\right)-\frac14}\
\Gamma_{\frac{b}{\sqrt2}}\left({\textstyle\frac{b}{\sqrt 2}}\right)
\frac{\Gamma_b(x)\Gamma_b\left(x+{\textstyle\frac{1}{2}}b\right)}{\Gamma_b\left({\textstyle\frac{1}{2}}b\right)\Gamma_b(b)}\ .
\end{equation}
The function
\begin{eqnarray*}
H(b,x;s)
& = &
2\zeta_b\left({\textstyle\frac{1}{2}}Q;s\right) - \zeta_b(x;s) - \zeta_b(Q-x;s)
\end{eqnarray*}
satisfies
\begin{eqnarray*}
\frac{\partial}{\partial s} H(b,x;s)\Big|_{s=0} & = & \log\Upsilon_b(x),
\hskip 1cm
H(b,x;0) \;  = \;  -\left({\textstyle\frac{1}{2}}Q-x\right)^2.
\end{eqnarray*}
Repeating for $H$ the previous calculation one obtains:
\[
H\left({\textstyle\frac{b}{\sqrt 2}},x\sqrt 2;s\right)
-
H\left({\textstyle\frac{b}{\sqrt 2}},{\textstyle\frac{b}{\sqrt 2}};s\right)
=
2^{-\frac{s}{2}}
\Big\{
H\left(b,x;s\right) +H\left(b,x+{\textstyle\frac{1}{2}}b;s\right) - H\left(b,{\textstyle\frac{1}{2}}b;s\right) - H(b,b;s)
\Big\}.
\]
This implies the FLNO shift formula:
\begin{equation}
\label{Upsilon:scaling}
\Upsilon_{\frac{b}{\sqrt2}}(x\sqrt{2})
=
2^{x\left(x-\frac{1}{b}-\frac12 b\right)+\frac12}\
\Upsilon_{\frac{b}{\sqrt2}}\left({\textstyle\frac{b}{\sqrt 2}}\right)
\frac{\Upsilon_b(x)\Upsilon_b\left(x+{\textstyle\frac{1}{2}}b\right)}{\Upsilon_b\left({\textstyle\frac{1}{2}}b\right)\Upsilon_b(b)}\ .
\end{equation}
Finally, replacing in (\ref{Upsilon:scaling}) $b \to b^{-1}$ one gets the relation:
\begin{equation}
\label{Upsilon:inverse:scaling}
\Upsilon_{b\sqrt 2}(x\sqrt{2})
=
2^{x\left(x-\frac{1}{2b}- b\right)+\frac12}\
\Upsilon_{b\sqrt 2}\left({\textstyle\frac{b^{-1}}{\sqrt 2}}\right)
\frac{\Upsilon_b(x)\Upsilon_b\left(x+{\textstyle\frac{1}{2}}b^{-1}\right)}{\Upsilon_b\left({\textstyle\frac{1}{2}}b^{-1}\right)\Upsilon_b(b^{-1})}\ .
\end{equation}

\end{document}